\documentclass{evn2004}
\usepackage{txfonts}
\usepackage{graphicx}

\def\ltsim  {\ifmmode\stackrel{<}{_{\sim}}\else$\stackrel{<}{_{\sim}}$\fi}
\def\gtsim  {\ifmmode\stackrel{>}{_{\sim}}\else$\stackrel{>}{_{\sim}}$\fi}
\def\farcs  {\hbox{$.\!\!^{\prime\prime}$}}

\def\kms    {\ifmmode {\rm km\,s}^{-1} \else km\,s$^{-1}$\fi}

\begin{document}

\title{OH in Messier 82}

\author{M.K. Argo\inst{1}, A. Pedlar\inst{1}, T.W.B. Muxlow\inst{1}, R.J. Beswick\inst{1}
, S. Aalto\inst{2}, K. Wills\inst{3}, R. Booth\inst{2}}
\institute{Jodrell Bank Observatory, University of Manchester, Macclesfield, Cheshire, SK11 9DL, UK
	\and
	Onsala Space Observatory, Chalmers University of Technology, S-439 92 Onsala, Sweden
	\and
	Dep. Physics and Astronomy, University of Sheffield, Sheffield, S7 3RH, UK}

\abstract{
Several new main line OH masers have been detected in the nearby starburst galaxy M82. Eight 
masers have been detected to 5$\sigma$, six of which are new detections.  Observations covering 
both the 1665 and 1667\,MHz lines with both the Very Large Array (VLA) and the Multi-Element 
Radio Linked Interferometer Network (MERLIN) have been used to accurately measure the positions 
and velocities of these features.  Following analysis of the data, another six objects below 
5$\sigma$, but with velocities consistent with the distribution inferred from the more certain 
detections, have been detected.  These are classified as possible detections.
}

\maketitle

\section{Introduction}

M82 is one of the closest, and therefore well studied, starburst galaxies and radio 
observations of this galaxy are numerous.  Extensive studies of the distribution and dynamics 
of neutral hydrogen have been made (e.g. Wills et al 2000, 2002) and these have been compared 
with the molcular gas as traced by CO emission (Shen \& Lo 1995).  The angular resolution of 
the CO observations is however comparatively low.  Observations of the transitions of the OH 
molecule can be made with the same instruments at similar resolution to the previous H{\sc i} 
studies and so provide a better comparison.

An OH maser was first detected in M82 using the Effelsberg telescope by Nguyen-Q-Rieu et al 
(1976) and two bright main line masers were detected by Weliachew et al (1984) using the VLA.  
In addition, Seaquist et al (1997) find six OH features at the 1612 and 1720\,MHz satellite 
lines.  H$_{2}$O masers were also detected by Baudry \& Brouillet (1996).

Recently the VLA has been used to probe the OH absorption across the central starburst region 
in M82.  As a by-product of this observation, eight masers have been detected to 5$\sigma$, six 
of which are new detections.  Another six objects are possible detections: these are features 
below 5$\sigma$ but with velocities consistent with the distribution inferred from the more 
certain detections.

\begin{figure}
\centering
\includegraphics[width=6cm,angle=-90]{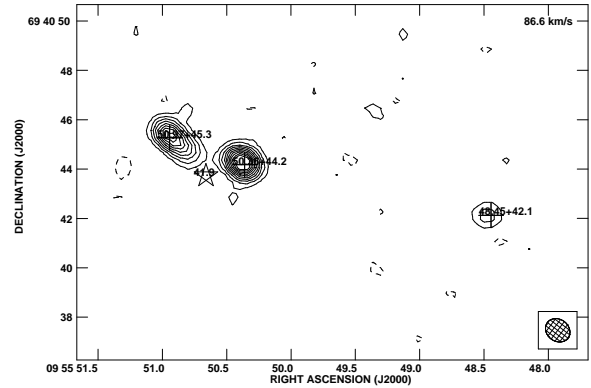}
\caption{Three of the central masers in a single channel from the VLA 2002 data set showing the 
extension of maser 50.97+45.3.  The position of the brightest continuum feature (the SNR 
41.9+57.5) is marked by a star.  Contours are at -1, 1, 2, 3, 4, 5, 6, 7, 8, 9, 10 $\times$ 
1.751 mJy/beam.}
\label{fig_extended}
\end{figure}

\begin{table}
\begin{center}
\begin{tabular}{ccc}
ID			& Velocity	& Nearest continuum\\
(J2000)			& (\kms)	& feature$^1$ (B1950)\\
\hline
48.45+42.1		& 116$\pm$9	& 39.68+55.6 (H{\sc ii}) \\
49.71+44.4		& 75$\pm$9	& 40.96+57.9 (H{\sc ii}) \\
50.36+44.2		& 80$\pm$9	& 41.64+57.9 (H{\sc ii}) \\
50.97+45.3		& 98$\pm$9	& 42.21+59.2 (H{\sc ii}) \\
51.23+44.5		& 116$\pm$9	& 42.48+58.4 (H{\sc ii}) \\
51.88+48.3		& 239$\pm$9	& -	\\
52.71+45.8		& 239$\pm$9	& 44.01+59.6 (SNR)      \\
53.62+50.1		& 362$\pm$9	& 44.93+63.9 (H{\sc ii}) \\
\end{tabular}
\caption{\label{maserTable}The eight definite maser detections.  The features are named 
according to their J2000 positions (relative to 09$^{\rm h}$55$^{\rm m}$ +69\degr40').  
Velocities are measured with respect to the line rest velocity and where a maser is detected in 
both lines the velocities are consistent between the two.  The nearest continuum features are 
from, and are labelled according to, the convention used in McDonald et al (2002).}
\end{center}
\end{table}

\begin{figure*}
\centering
\includegraphics[width=11cm,angle=-90]{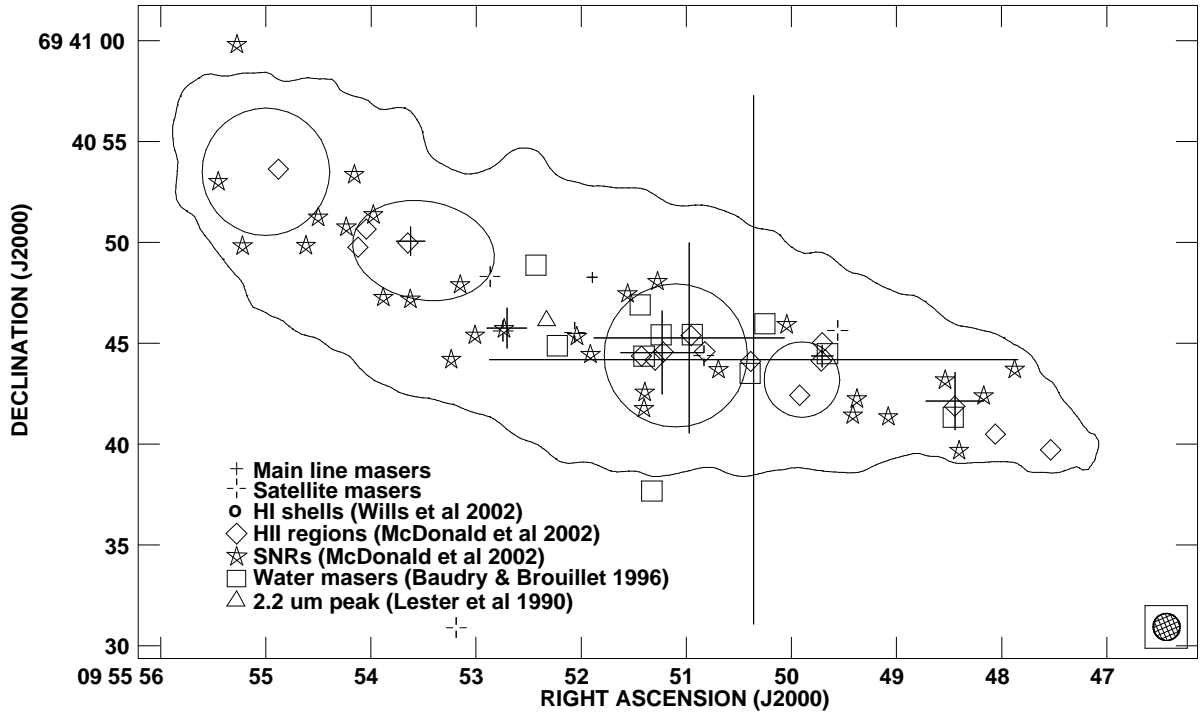}
\caption{The location of the main line masers in M82 with other features shown for comparison.  
The solid line is a 1$\sigma$ contour of the continuum emission from the VLA data cube.  The 
masers reported here are shown as crosses with symbol size scaled relative to the peak 
intensity measured in the VLA data set.}
\label{fig_all}
\end{figure*}

\section{Observations}

The VLA in A configuration was used in April and May 2002 to observe M82 with a bandwidth of 
6.25\,MHz over 64 channels with the result that both main line transitions were observed in the 
same data set, with a velocity resolution of $\sim$18\,km\,s$^{-1}$.  After 
calibrating these data using standard VLA techniques, the image was continuum subtracted using 
the line free channels in order to separate out the line component.  Eight maser spots were 
identified in this data set above a limit of 5$\sigma$, six of these are new detections.

A MERLIN observation of M82, taken in 1995, has also been examined.  This data set has a higher 
spatial resolution than the more recent VLA observation (0\farcs1 to 0\farcs2 compared to 
$\sim$1\farcs4), although as the bandwidth was 8\,MHz over 64 channels the spectral resolution 
is $\sim$22\,km\,s$^{-1}$.  A higher spectral resolution VLA data set from 19 
November 1988, covering only the 1665\,MHz line, has also been examined.

In February 2004 the EVN was also used to observe OH in M82 at high spectral and spatial 
resolution.  This data set is yet to be analysed.

\section{Results}

One of the masers (50.97+45.3) is clearly extended in the low resolution VLA data set from 2002 
(Fig. \ref{fig_extended}) and appears to consist of at least three velocity components in the 
higher spectral resolution VLA data set from 1988.

Figure \ref{fig_all} (Argo et al, in prep.) shows the positions of the masers (crosses) 
detected in the VLA and MERLIN observations.  Also plotted in the same figure are the known 
supernova remnants and H{\sc ii} regions (stars and diamonds respectively, 
McDonald et al 2002), H{\sc i} shells (ellipses, Wills et al 2002), satellite line masers 
(broken crosses, Seaquist et al 1997), water masers (squares, Baudry \& Brouillet 1996) and the 
2.2$\mu$m peak (Lester et al 1990).

The two brightest masers in these data are coincident with those reported by Weliachew et al 
(1984).  Four are coincident with the water masers of Baudry \& Brouillet (1996) and the 
velocities measured are also consistent to within the errors between these two observations.

Most of the definite detections are observed in both main lines and are coincident with 
either known supernova remnants or H{\sc ii} regions.  In general, masers which are brighter at 
1667\,MHz are associated with HII regions, while those that are brighter at 1665\,MHz are 
generally coincident with SNRs.

Of the possible detections, only one is coincident with a known continuum feature.

\section{Conclusions}

It is likely that there are more main line OH masers in M82 but due to the depth of absorption 
and the low velocity resolution faint or narrow masers could be buried to the extent that they 
are undetectable in these observations.

\end{document}